# A Stellar Audit: The Computation of Encounter Rates for 47 Tucanae and $\omega$ Centauri

Melvyn B. Davies *
*Theoretical Astrophysics, Caltech, Pasadena CA 91125.*

Willy Benz
*Steward Observatory, University of Arizona, Tucson AZ 85721.*

**ABSTRACT**

Using King-Mitchie Models, we compute encounter rates between the various stellar species in the globular clusters $\omega$ Cen, and 47 Tuc. We also compute event rates for encounters between single stars and a population of primordial binaries. Using these rates, and what we have learnt from hydrodynamical simulations of encounters performed earlier, we compute the production rates of objects such as low-mass X-ray binaries (LMXBs), smothered neutron stars and blue stragglers (massive main-sequence stars).

If 10% of the stars are contained in primordial binaries, the production rate of interesting objects from encounters involving these binaries is as large as that from encounters between single stars. For example, encounters involving binaries produce a significant number of blue stragglers in both globular cluster models. The number of smothered neutron stars may exceed the number of low-mass X-ray binaries (LMXBs) by a factor of 5-20, which may help explain why millisecond pulsars are observed to outnumber LMXBs in globular clusters.

## INTRODUCTION

The products of stellar encounters are proving to be a powerful tool for investigating the life-history of globular clusters. In this paper, we employ the cross sections for the formation of detached or merged systems through two-body encounters calculated from hydrodynamic simulations of stellar encounters (Benz & Hills 1987, 1991; Davies, Benz & Hills 1991, 1992, hereafter DBH91, DBH92) to compute the encounter rates in the globular clusters $\omega$ Cen, and 47 Tuc using the King-Mitchie models fitted by Meylan (1987, 1989) for the distributions of the various stellar species. We then consider the role of encounters involving primordial binaries, using the results obtained by Davies (1995). In applying what we learnt from the encounters simulated with our SPH code, we hope to address questions concerning the origins of objects observed in globular clusters over a broad range of the electromagnetic spectrum: including radio objects such as millisecond pulsars (MSPs), objects visible in the optical such as blue stragglers and blue subdwarfs, and X-ray binaries.

A clear illustration of our current fragmentary understanding of globular clusters is the current debate over the origin of MSPs. Under the standard model, MSPs are produced in low-mass X-ray binaries (LMXBs) where the neutron star is spun-up as material is accreted from the Roche-filled companion. However observations seem to suggest that there are far more MSPs than LMXBs which, given their comparable expected lifetimes, poses a problem for the standard model (Kulkarni et al. 1990). MSPs continue to be discovered in ever-increasing numbers (Manchester et al. 1991). This embarrassing profusion of MSPs coupled with the relative sparsity of LMXBs would seem to represent the nemesis of the standard model.

Recent models of LMXB evolution have suggested that these systems have radically shorter lives than previously thought when the effects of the flux from the pulsar on the Roche-filled companion are taken into account (Frank, King & Lasota 1992), however these effects may be less than at fisrt thought (see for example Ritter 1994). Although a shortening of LMXB lifetimes may offer a part of the solution, recent observational evidence further befogs an already murky picture of the processes at play in globular clusters. Johnston, Kulkarni & Goss (1991) report a failure to detect *any* MSPs in Liller 1, a rich cluster that is believed to have one of the densest cores of all globular clusters in the Galaxy. We therefore must conclude that the production rate of MSPs, through whatever process is dominant in clus-ters, is not a simple function of cluster core density. Johnson, Phinney & Kulkarni (1991) concluded from an analysis of all published surveys of globular clusters that the number of pulsars in a cluster, $N_p$, is best fitted by the relation $N_p \propto M_c \rho_c^{1/2}$. Phinney & Kulkarni (1994) claim that this power-law dependence can be explained by considering tidal interactions with primordial binaries. Alternatively, a large fraction of the observed MSPs may be formed via a mechanism know as Accretion Induced Collapse (AIC), as suggested by Grindlay (1987). In AIC, a massive white dwarf surpasses the Chandrasekhar limit by accreting material and collapses to form a neutron star. Whether the star can avoid complete disruption as a supernova is currently a matter of very active debate within the astronomical community (see for example Verbunt et al. 1989).

* Current Address: Institute of Astronomy, Madingley Road, Cambridge CB3 0HA

In reviewing the products of various encounters, and examining the production rates of the various stellar exotica we see in the simulations, we remain particularly vigilant for any objects that may permit the accretion of material onto the surface of a neutron star whilst conspiring to obscure any X-rays produced through the accretion of matter. Cases where a smaller amount of material is accreted may well be important in producing MSPs of periods $\geq 10$ms.

Blue stragglers are positioned on the upper end of the main-sequence beyond the present day turn-off mass. They have been observed in many globular clusters, including: M3 (Sandage 1953), M71 (Arp & Hartwick 1971), NGC 6352 (Hartwick and Hesser 1972), and NGC 5466 (Nemec & Harris 1987). Severe crowding problems have, until recently, inhibited searches for blue stragglers in cores of dense clusters. A study by Paresce et al. (1991) of the core of 47 Tuc, using the Faint Object Camera (FOC) on the HST, revealed a high density of blue stragglers. Also recently, a study by Auriere et al. (1990), using the ESO NTT telescope, revealed a population of blue stragglers in the center of the post-collapse cluster NGC 6397. The current consensus in the astronomical community attributes blue stragglers to one of two possible mechanisms. They could have formed from collisions between main-sequence stars, as such collisions lead to the merger of the two stars and thorough mixing of the material, resetting the "stellar clock" (Benz & Hills 1987). Alternatively, such systems may have formed through the coalescence of two main-sequence stars in an extremely hard binary, or may still be a contact binary today (see for example Margon & Cannon 1989, and Mateo et al. 1990). The former mechanism may be more efficient at dredging up material to the surface of the merged object, hence by deducing the photospheric abundance of helium or heavier elements, we may be able to infer which of these two mechanisms is favored in globular clusters. In either mechanism, we would expect blue stragglers to have masses larger than the majority of the stellar components of a globular and would therefore sink in the gravitational potential of the core (in a timescale comparable to the local relaxation time). This expectation seems consistent with observations, with distributions being observed to be strongly centrally peaked and estimates for blue-straggler masses falling in the range $1.0 - 1.6 M_\odot$. If blue stragglers are indeed formed by the collision of two main-sequence stars, these objects will provide us with important information concerning the evolutionary history of globular clusters.

Color gradients have been observed in post-collapse clusters (Djorgovski et al., 1991 and references contained therein), with the centers being somewhat bluer. It is unlikely that cataclysmic variables (CVs) in the cluster cores are the cause of the color gradient, as very little H$\alpha$ emission (CVs are strong in H$\alpha$) has been seen, at least in the cluster M15 (Cederbloom et al. 1992). Alternatively, the color gradient may be due to a population of blue subdwarfs. These stars are positioned on the blue edge of the horizontal branch in a color-magnitude diagram, and are thought to consist of $0.45 M_\odot$ helium-burning cores surrounded by an envelope of mass $\leq 0.05 M_\odot$. Several mechanisms have been suggested for their production. Mass loss either during the horizontal branch phase (Demarque & Eder 1985), or accompanying the helium flash (Heber et al. 1984) have been suggested. Alternatively, Iben & Tutukov (1984, 1985, 1986, 1987) believe that such systems are produced by the merger of two helium white dwarfs. Such stars in a tight binary will spiral in and merge as angular momentum is lost though gravitational radiation. Mengel, Sweigart & Demarque (1976) suggested that two stars, of mass $0.8 M_\odot$ and $0.78 M_\odot$, in a binary of initial separation $0.7 - 0.8$ AU, may produce a blue subdwarf from the primary as enough of its envelope would be lost in mass-transfer to the secondary, as the former evolved up the giant branch. However, if the initial binary separation were smaller, mass transfer would remove the primary envelope completely, leaving a He white dwarf. For greater initial separations, the primary would fail to fill its Roche Lobe on its ascent up the red giant branch, and would produce a normal horizontal branch star. One uncertainty in Mengel's calculations concerns the nature of mass-transfer. He assumes the transfer would be conservative. In his models, mass transfer occurs in a stable fashion, with the separation of the two stars increasing as material is accreted by the secondary. If, however, material – and angular momentum – are lost from the system, the two stars may well spiral together. Djorgovski et al. (1991) has suggested placing an $0.8 M_\odot$ star in a hard binary with a low-mass He white dwarf. As the primary fills its Roche Lobe, unstable mass transfer will ensue, leaving a common envelope system. If this occurs as the primary is ascending the red giant branch, the core of the star will in effect also be a low-mass He white dwarf. The two white dwarfs will spiral together as the envelope is ejected, ultimately leaving an extremely hard binary, whose two components will spiral together and merge as described by Iben and Tutukov.

A recent study of the core of M15 by Murphy et al. (1991) demonstrated the existence of sources of emission in the cores of the Ca II H and K absorption lines. Such emission is produced by chromospheric activity which has been associated with a magnetic field produced by a high rotation rate (Hartman & Noyes 1987); alternatively it has been suggested that a pulsation instability produces shocks that lead to emission (Dupree et al. 1990), or that the emission lines are simply indicative of low metallicity (Dupree & Whitney 1991). An obvious way to produce a rapidly-rotating star is to place it in a binary and synchronize stellar rotation with orbital motion. Over the evolution of a globular cluster, primordial binaries will either be broken up or will harden. Of those surviving, the average spacing is expected to lie between 0.2 and 0.5 AU (Hut et al. 1991). In fact, a binary produced through the tidal capture of a compact object by a red giant is also likely to have a similar separation

(DBH91, DBH92). The periods of many of these binaries would be short enough to produce the requisite magnetic activity giving the observed Ca II H and K lines. Thus it has been suggested that the presence of emission in the cores of the Ca II H and K lines could be used as a diagnostic tool for inferring the distribution and abundance of binaries in globular clusters. This tool could certainly be employed in observing globular cluster cores with the HST.

This paper is organized as follows. In §2, we review the King-Mitchie model for globular clusters and apply it to the two clusters $\omega$ Cen, and 47 Tuc. In §3, we compute the cross sections for binary formation and mergers between two stars. The production rates from encounters between primordial binaries and single stars are computed for the two globular clusters in §4. We compare the binary/single and single star/single star rates in §5.

## 2 APPLYING KING-MITCHIE MODELS TO $\omega$ CEN, AND 47 TUC

We fitted King-Mitchie models to two particular globular clusters: $\omega$ Cen, and 47 Tuc, as has be done already by Verbunt & Meylan (1988). Following Verbunt & Meylan (1988), we model the stellar population by some power law, where the number of stars having a mass between $m$ and $m+d\log m$ is given by $dN(m) \propto m^{-x} d\log m$. The stars are binned by mass into different classes $j$ and the stellar distributions are computed using a distribution function having a lowered Maxwellian energy dependence (King 1966), with an anisotropy term taking the form $\exp(-\beta J^2)$ (Mitchie 1962). For a further discussion of King-Mitchie models, see Davies (1995).

Using the globular characteristics for $\omega$ Cen, and 47 Tuc given by Meylan (1987, 1989), we reproduce his Model 7 for $\omega$ Cen, Model 13 for 47 Tuc. The properties of the two globulars are given in Table 1. Inspection of this table reveals that the two globular clusters considered here are quite different. 47 Tuc has a far more compact core with a much higher central density. In addition, their stellar populations are quite different. For $\omega$ Cen, the stellar population is fit by a single power law, $dN(m) \propto m^{-x} d\log m$, where $x = 1.0$. In the case of 47 Tuc, two different values of $x$ are employed. For stars beneath the turn-off, $x = x_{\rm ms} = 0.2$. For stars initially more massive than the present day turn-off ($0.88 M_\odot$), $x = 3.0$. Using these power laws the initial population (all main-sequence stars) is computed for the two globular clusters. Stars of mass $\geq 6 M_\odot$ are assumed to produce neutron stars of mass $1.4 M_\odot$ (mass bin 12). Those of mass, $2.5 M_\odot \leq M_\star \leq 6 M_\odot$ produce white dwarfs which are placed in mass bin 11, whilst those in the ranges, $1.5 \leq M_\star \leq 2.5$ and $M_{\rm turn-off} \leq M_\star \leq 1.5$ are assumed to produce white dwarfs which are placed in mass bins 9 and 10 respectively. The last two mass bins have the same average masses as the two most massive main-sequence star bins, i.e. mass bins 7 and 8. The fraction of stars contained in each mass bin is shown for the two globular clusters in Table 2 together with the average mass of stars in each bin, $M_{\rm av}$. We immediately notice that the mass fraction of heavy remnants, i.e. stars in mass bins 11 and 12, is greatly depleted in 47 Tuc because of the steep power law employed for the initial population more massive than the present-day turn-off mass. Instead,

**Table 1.** Properties of the globular cluster models for $\omega$ Cen and 47 Tuc.

| Property | $\omega$ Cen | 47 Tuc |
|---|---|---|
| $\rho_0$ ($M_\odot/{\rm pc}^3$) | 4.0E+3 | 5.1E+4 |
| $M_{\rm tot}$ ($10^6 M_\odot$) | 3.34 | 1.06 |
| $r_c$ (pc) | 3.4 | 0.59 |
| $r_t$ (pc) | 74.0 | 69.0 |
| $r_a/r_c$ | 3.0 | 40. |
| $v_0$ (km/s) | 16.8 | 10.3 |
| $t_{\rm rh}$ ($10^9$ years) | 16.0 | 8.7 |
| $t_{\rm rc}$ ($10^6$ years) | 490.0 | 12.0 |

$\rho_0$ is the central density, $M_{\rm tot}$ is the total mass of the cluster, $r_c$ the core radius, $r_t$ the tidal radius, $r_a$ the anisotropy radius, $v_0$ the central velocity dispersion, $t_{\rm rh}$ the half-mass relaxation timescale, and $t_{\rm rc}$ the central relaxation timescale.

this globular cluster contains more main-sequence stars close to the turn-off mass. These differences will manifest themselves when we compute the two-body encounter rates for these three models in Section 5.

Using the globular cluster parameters given in Table 1, together with the stellar populations given in Table 2, the spatial distributions for each mass bin, together with their velocity dispersions, was obtained.

To obtain the total stellar population in the core, we integrated the number densities of the various stellar species out to a scale radius. The stellar population in the core of a cluster differs from the IMF of the whole cluster; mass segregation results in the core containing more massive stars. This effect is more pronounced in more collapsed clusters (i.e. clusters with a larger value of $W_0$), as can be seen in Table 2. The mass fraction contained in the various mass bins in the cores of the two clusters is given in Table 2, together with the mass fractions for the whole cluster. For $\omega$ Cen mass bin 12 makes the greatest contribution to the central density, followed by mass bins 8-10. In 47 Tuc, the cluster is sufficiently devoid of neutron stars that their contribution is more than an order of magnitude smaller than that from the bins containing the white dwarfs and the most massive main-sequence stars. It will come as no surprise when we see this observation repeated when we compute the two-body encounter rates and find that in 47 Tuc encounters between main-sequence stars and white dwarfs or other main-sequence stars are more common than encounters between main-sequence stars and neutron stars.

## 3 ENCOUNTERS BETWEEN TWO SINGLE STARS

Before we consider encounters between binaries and single stars, we consider the rate of encounters between two single stars, given the core models in Table 2. The rate of encounters between two stellar species, 1 and 2, passing within some minimum distance, $R_{\rm min}$ is given by

$$\Gamma_{12} = 6 \times 10^{-9} \frac{n_1}{10^4 \text{ pc}^{-3}} \frac{n_2}{10^4 \text{ pc}^{-3}} \frac{M_1 + M_2}{M_\odot} \\ \times \frac{R_{\rm min}}{R_\odot} \frac{10 \text{kms}^{-1}}{V_\infty} \text{ yr}^{-1} \text{pc}^{-3} \quad (1)$$

**Table 2.** The mass fraction contained in each mass bin for the models of $\omega$ Cen and 47 Tuc.

| mass class | nature | $\omega$ Cen $M_{av}$ | MF | MF$_c$ | 47 Tuc $M_{av}$ | MF | MF$_c$ |
|---|---|---|---|---|---|---|---|
| 1 | MS | 0.14 | 0.080 | 0.025 | 0.14 | 0.040 | 0.002 |
| 2 | MS | 0.18 | 0.080 | 0.028 | 0.18 | 0.050 | 0.004 |
| 3 | MS | 0.22 | 0.080 | 0.030 | 0.23 | 0.060 | 0.005 |
| 4 | MS | 0.28 | 0.080 | 0.034 | 0.28 | 0.080 | 0.008 |
| 5 | MS | 0.35 | 0.080 | 0.040 | 0.36 | 0.090 | 0.015 |
| 6 | MS | 0.44 | 0.080 | 0.049 | 0.45 | 0.110 | 0.032 |
| 7 | MS | 0.58 | 0.080 | 0.066 | 0.59 | 0.160 | 0.115 |
| 8 | MS | 0.73 | 0.080 | 0.089 | 0.77 | 0.210 | 0.467 |
| 9 | WD | 0.58 | 0.120 | 0.098 | 0.59 | 0.150 | 0.107 |
| 10 | WD | 0.73 | 0.070 | 0.078 | 0.77 | 0.040 | 0.089 |
| 11 | WD | 1.10 | 0.080 | 0.165 | 1.09 | 0.010 | 0.130 |
| 12 | NS | 1.40 | 0.090 | 0.299 | 1.40 | 0.001 | 0.025 |

For each model, we list the average mass of each mass bin ($M_{av}$), the mass fraction for the whole cluster (MF) and the mass fraction for the stars in the core (MF$_c$).

**Table 3.** The production rates of various stellar systems.

| Producing | $\omega$ Cen | 47 Tuc |
|---|---|---|
| MS–MS | 1.19 | 4.50 |
| WD–MS | 1.64 | 6.13 |
| NS–MS | 1.36 | 0.33 |
| MS–RG | 0.22 | 1.78 |
| WD–RG | 0.18 | 1.75 |
| NS–RG | 0.15 | 0.09 |

The different systems listed are: main-sequence/main-sequence star (MS–MS), white dwarf/main-sequence star (WD–MS), and neutron star/main-sequence star (NS–MS) systems, (per $10^8$ years) by encounters between single stars for the globular cluster models for $\omega$ Cen and 47 Tuc. Also listed are the production rates for encounters involving red giants (RGs).

where $n_1$ and $n_2$ are the number densities of the two species, and $V_\infty$ is given by (Verbunt & Meylan 1988)

$$\left\langle \frac{1}{V_\infty} \right\rangle = \sqrt{\frac{6}{\pi} \frac{1}{V_1^2 + V_2^2}} \qquad (2)$$

We use an analytic expression for $R_{min}$ suggested by Benz and Hills (1991):

$$\frac{R_{min}}{R_1} = 3.28 + 0.45 \log\left(\frac{M_2}{M_1}\right) + 0.22 \log\left(\frac{M_1}{M_\odot} \cdot \frac{R_\odot}{R_1}\right) \qquad (3)$$

We thus calculate the encounter rates between the various stellar species for the models listed in Table 2; the results being listed in Table 3, with production rates being given in units of encounters/$10^8$ years. We assume here that all close encounters between two main-sequence stars that become bound produce blue stragglers if their total mass is larger than the turn-off mass. We also listed the rates for encounters involving red giants (assumed to be half the rate for the most massive main-sequence stars). We can clearly see that the core of 47 Tuc contains more massive main-sequence stars than the core of $\omega$ Cen, as the event rate for red giants is $\sim 1/4$ that of the total in 47 Tuc, compared to a figure of $\sim 1/6$ in $\omega$ Cen. Not shown in Table 3 is the distribution of rates over mass. Again we see the effect of the flatter IMF in 47 Tuc. A much larger fraction of the encounters in 47 Tuc involve the most massive main-sequence stars compared to those in $\omega$ Cen.

There has been much debate concerning whether binaries can be formed via tidal capture. Here we simply list all close encounters between main-sequence stars and white dwarfs and neutron stars. Some fraction of these encounters may lead to binaries rather than a smothered compact object. Simulations suggest that less than half the encounters produce binaries (DBH92). If binaries are produced, they will quickly circularise to separations $\sim$ twice that of the initial separation at closest approach (*i.e.* $\sim 6R_{ms}$). Such tight binaries are likely to be brought into contact through angular momentum loss from a magnetized wind from the main-sequence star or red giant. Once brought into contact, the subsequent evolution of a binary is a function of the ratio of the masses of the two stars. Mass transfer will be stable if the size of the main-sequence star's Roche Lobe increases faster than the radius of the star, thus a necessary but not *sufficient* condition of stable mass transfer is that the separation of the two stars increases. For conservative mass transfer, this requires that the star filling its Roche Lobe be less massive than its companion. In fact it has been suggested that we require a larger value of the mass ratio to produce stable mass transfer (Verbunt & Rappaport 1988, Bailyn, Grindlay & Garcia 1990). Thus a fraction of the detached binaries containing compact objects will not produce LMXBs or CVs. Rather these systems will produce a merged object after an extremely brief period of unstable mass transfer where the compact object will become smothered by the material from the main-sequence star. Such systems will be similar to those seen in closer encounters and physical collisions.

Our simulations of encounters between neutron stars and main-sequence stars showed that for physical collisions and close encounters, the main-sequence star forms a highly-flattened envelope around the neutron star. The fate of such a system is currently unclear as no one has modelled the dynamical evolution of a system with a neutron star being smothered by an envelope. For more massive envelopes ($M_{env} \geq 4M_\odot$), a stable envelope distribution has been calculated, using a stellar structure code. In these so called Thorne-Zytkow objects, matter from the envelope will accrete onto the neutron star until it becomes sufficiently massive to collapse to a black hole. The accretion of only $0.025 M_\odot$ of material may be sufficient to spin-up the neutron star to a period of $\simeq 10$ ms, so perhaps *some* of the MSPs observed in globulars have been formed from these smothered-neutron-star systems. However it has been suggested that lower mass envelopes are susceptible to instabilities, shortening their life, and potentially decreasing the mass accreted below a level of importance. *Perhaps* enough material accretes onto the neutron star in the smothered systems, whilst the envelope itself conspires to obscure the X-rays until in the *final* phases of the accretion either the envelope becomes unstable or is blown off, thus leading to a X-ray lifetime for the object much shorter than that of an LMXB. Thus we will later be noting the relative number of smothered neutron star systems to clean binaries we expect will produce LMXBs.

In DBH91, we concluded that a close encounter between a red giant and a white dwarf would leave, after a common envelope phase, a white dwarf/white dwarf binary that would spiral-in because of angular momentum loss by gravitational radiation until the less-massive component filled its Roche lobe. This process could potentially be unstable and result in the formation of a thick disk around the more massive white dwarf (Benz et al. 1989). It has been suggested that such an object would produce a core-helium burning star visible as a subdwarf B star if the two compact objects are He white dwarfs. In addition, we concluded that a close encounter between a red giant and a neutron star will produce a similar system with the more massive white dwarf replaced by the neutron star (DBH92). Again unstable transfer of the white dwarf onto the neutron star would produce a thick disk of material. In this case, there is the potential for the neutron star to accrete enough material to spin it up to millisecond periods before the remainder of the disk is disrupted without requiring any instabilities to flatten the object. In such a way it may be possible to produce a MSP without first passing through a long-lived X-ray phase. An example of a system possibly produced by a collision between a neutron star and a red giant is the 11 minute X-ray binary 4U 1820-30 (Bailyn & Grindlay 1987, Verbunt 1987). In this system a white dwarf appears to be transferring mass in a *stable fashion* to its neutron star companion. Bailyn and Grindlay conclude that the mass of the white dwarf is $\leq 0.09 M_\odot$, hence if this system was produced through an encounter between a red giant and a neutron star, a common envelope phase must have begun when the red giant had not evolved very far up the giant branch. In other words, the neutron star must have struck a less-evolved red giant than the one considered in DBH92. Alternatively, Bailyn and Grindlay suggest that the neutron star first encountered a main-sequence star, forming a detached binary, later incurring mass-transfer when the latter star evolved off the main sequence.

## 4 ENCOUNTERS BETWEEN BINARIES AND SINGLE STARS

We now consider encounters between primordial binaries and single stars. Owing to their larger cross section, such encounters may be important even if their relative fraction is low. Our studies (Davies, Benz & Hills 1993, 1994) of encounters between binaries and single stars show three main outcomes: the incoming star replaces one of the original components forming a new, detached, binary in a so-called *clean exchange*, the incoming star simply hardens the binary without any exchange occuring – a so-called fly-by, or a merger between two of the stars occurs. In a companion work (Davies 1995), we described how we used a series of three-body simulations to produce expressions for these three outcomes; namely the exchange cross section, $\sigma_{cx}$; the fly-by encounter, $\sigma_{fb}$; and the cross section to form a merged object, $\sigma_{mb}$. The last outcome bifurcates; the merged object may be bound to the third star, producing a so-called *merged binary* (MB), or it may be unbound, a event that has been dubbed a *scattering-induced merger* (SIM).

We use the cross sections discussed above to simulate the evolution of a population of binaries in the cores of $\omega$ Cen and 47 Tuc. We treat each of the cores of the two clusters as a static sphere containing a population of stars calculated by integrating the number densities of the various stellar species out to a scale radius. In practice, the background of single stars will also be evolving, though the core may be supported from collapse by the energy released from the binaries. We imagine binaries diffusing into the core, encountering a single star (or perhaps many single stars) and producing some new binary/object. We neglect 2+1 encounters occuring outside of the core (this is not unreasonable as only a small fraction encounters will occur outside of the core). For a given binary, we compute its cross sections for interactions with all the different mass classes of single stars, select one at random (weighted by the cross sections) thus producing a new object. If the new system is a binary that is not ejected from the cluster, we repeat the process, thus simulating subsequent encounters. If the new binary is removed from the core, but not from the cluster, we estimate a timescale for the binary to return to the core and reinsert it into the core after the computed delay. We refer the reader to Davies (1995) for a more detailed discussion of the computation of return times.

We considered subsequent encounters involving the binary until one of the following occurred: 1) a merged object was produced in a scattering-induced merger (SIM), 2) the binary was deemed to be likely to be broken up through an encounter with a third star, 3) a single merged object (SMO) was produced, 4) the binary was ejected from the cluster, or 5) the binary was deemed to have come into contact via inspiral from either a magnetized wind or gravitational radiation. Additionally, in some runs we halted the evolution of the binaries when a smothered neutron star was produced, as the subsequent evolution of such an object is currently unclear.

There are clearly a great number of parameters we can vary; the cluster model, described by its IMF, and concentration, the population of binaries we feed into the cluster core, and their distribution of injection times. We also have to decide how long to evolve the cluster binaries for. Finally, we have to consider two stellar evolution issues; namely common envelope phases, and the treatment of smothered neutron stars. If one of the components in a binary expands to form a red giant it may engulf its companion to form a common envelope system. In such a system, the core of the red giant and its companion will spiral together as the envelope of gas is ejected, the change in the binding energy of the two objects ($\Delta E_g$) being equated to the binding energy of the envelope ($E_{env}$), within some efficiency, $\alpha_{ce}$, where $E_{env} = \alpha_{ce} \Delta E_g$. As the subsequent evolution of smothered neutron stars is uncertain, we treat them in two ways, either terminating the evolution of a particular binary as soon as a neutron star is engulfed, or treating them in the same manner as smothered white dwarfs assuming the envelope expands to red giant proportions.

Rather than explore the full volume of parameter space here, we show a restricted set, changing only one of the above parameters at a time in an attempt to show trends.

### 4.1 Encounters in $\omega$ Cen

We consider first encounters involving binaries in $\omega$ Cen. In run 1, we injected 1000 binaries (each containing two

**Table 4.** The outcomes of the evolution of binaries in various cluster models.

| Cluster | Run | $n_{\rm bin}$ | $\alpha_{ce}$ | BP | $t_{\rm evol}$ | $t_{\rm inj}$ | BS | TDS | contact binaries ||||| MGRI |||||
|---|---|---|---|---|---|---|---|---|---|---|---|---|---|---|---|---|---|
| | | | | | | | | | XB | CV | SN | SW | BS | NW | $\rm WD^2_a$ | $\rm WD^2_b$ | $\rm NS^2$ |
| $\Omega$ Cen | 1 | 1000 | 0.4 | A | $\infty$ | 0.0 | 234 | 136 | 104 | 72 | 6 | 83 | 75 | 24 | 11 | 0 | – |
| | 2 | 1000 | 0.4 | A | 15.0 | 0.0 | 101 | 35 | 9 | 6 | 1 | 10 | 26 | 0 | 0 | 0 | – |
| | 3 | 5000 | 0.4 | A | 15.0 | 15.0 | 312 | 66 | 15 | 23 | 1 | 34 | 43 | 0 | 0 | 0 | – |
| | 4 | 5000 | 0.4 | C | 15.0 | 15.0 | 230 | 61 | 12 | 9 | 0 | 17 | 46 | 1 | 0 | 0 | – |
| | 5 | 5000 | 0.4 | D | 15.0 | 15.0 | 161 | 34 | 13 | 14 | 0 | 10 | 30 | 2 | 1 | 0 | – |
| | 6 | 5000 | 0.4 | D | 15.0 | 15.0 | 146 | – | 16 | 16 | 0 | 8 | 24 | 0 | 0 | 0 | 0 |
| | 7 | 5000 | 0.1 | D | 15.0 | 15.0 | 141 | – | 12 | 7 | 0 | 3 | 27 | 1 | 2 | 0 | 0 |
| | 8 | 5000 | 0.1 | D | $\infty$ | 15.0 | 939 | – | 371 | 183 | 53 | 168 | 558 | 378 | 88 | 10 | 223 |
| 47 Tuc | 9 | 1000 | 0.4 | B | $\infty$ | 0.0 | 491 | 34 | 8 | 89 | 4 | 99 | 114 | 11 | 67 | 5 | – |
| | 10 | 1000 | 0.4 | B | 15.0 | 0.0 | 488 | 32 | 9 | 63 | 1 | 100 | 116 | 5 | 57 | 3 | – |
| | 11 | 1000 | 0.4 | B | 15.0 | 15.0 | 442 | 19 | 3 | 50 | 1 | 74 | 65 | 4 | 34 | 1 | – |
| | 12 | 1000 | 0.4 | C | 15.0 | 15.0 | 384 | 20 | 4 | 54 | 0 | 53 | 75 | 2 | 22 | 3 | – |
| | 13 | 1000 | 0.4 | D | 15.0 | 15.0 | 352 | 22 | 3 | 58 | 1 | 39 | 77 | 2 | 26 | 0 | – |
| | 14 | 1000 | 0.4 | D | 15.0 | 15.0 | 335 | – | 7 | 54 | 0 | 54 | 58 | 4 | 29 | 2 | 1 |
| | 15 | 1000 | 0.1 | D | 15.0 | 15.0 | 382 | – | 2 | 34 | 1 | 35 | 53 | 10 | 35 | 0 | 1 |
| | 16 | 1000 | 0.1 | D | $\infty$ | 15.0 | 399 | – | 6 | 87 | 1 | 66 | 102 | 24 | 102 | 5 | 1 |

$W_0$ is the depth of the cluster potential (in units of the central velocity dispersion). $\alpha$ is the exponent of the power-law IMF. $\alpha_{ce}$ is the efficiency in the common envelope stage. BP is the binary population: in type A, the initial binaries always contain two $0.73 M_\odot$ main-sequence stars, in type B they contain two $0.77 M_\odot$ main-sequence stars, in type C they contain two main-sequence stars drawn independently from the IMF, with the extra criterian that their total mass $> 1 M_\odot$, types D and E are identical to type C except that the minimum total mass is $0.6 M_\odot$ and $0.7 M_\odot$, respectively. $t_{\rm evol}$ is the total time the binaries were allowed to evolve (in Gyr), with the binaries being initially injected into the core at a time drawn randomly between zero and $t_{\rm inj}$, given in (Gyr). BS is the number of blue stragglers produced when two main-sequence stars collided during an encounter. TZO is the number of smothered neutron stars produced (when this was one of the criteria for terminating the evolution of the binary). Contact binaries are those systems brought into contact via angular momentum loss by magnetic winds. This outcome is further subdivided into: X-ray binaries (XB) when the system contains a neutron star more massive then its main-sequence companion, cataclysmic variables (CV) when the system contains a white dwarf more massive than its main-sequence companion, smothered neutron stars (SN) and smothered white dwarfs (SW) when the compact obejects are less massive than their companions, and blue stragglers (BS) when both components are main-sequence stars.

Systems brought into contact via gravitational radiation are labelled MGRI (mergers through gravitational radiation inspiral). This outcome is further subdivided into: neutron star/white dwarf mergers (NW), white dwarf/white dwarf mergers ($\rm WD^2$), and neutron star/neutron star mergers ($\rm NS^2$). $\rm WD^2$ are further subdivided into those where the total mass exceeds the Chandrasekhar mass ($\rm WD^2_a$) and those of smaller total mass ($\rm WD^2_b$).

$0.73 M_\odot$ main-sequence stars) into the core and let them evolve until one of the criteria descibed above was met. The initial binary separations were chosen randomly with equal probability in $\log d$ from the range $100 R_\odot < d < 533 R_\odot$ (the upper bound is set from our requirement that the initial binaries be resilient to break up from passing field stars, a condition which requires $M_1 M_2/d \sim 0.001$). In this run, we stop the evolution of a binary as soon as a smothered neutron star is produced, labelling the outcome as a thick-disk system (TDS). We also adopted $\alpha_{ce} = 0.4$. In Figure 2 we plot the final outcomes of the 1000 binaries as a function of when the outcomes occurred. The first thing to note from this plot is that a fraction of the binaries are not resolved until much later than 15Gyr after they have been inserted into the cluster core. In fact, some binaries would still exist after $\sim 150$Gyr. In other words, only a small fraction of the binaries in the core of $\omega$ Cen will have suffered many encounters. Figure 2 also tells us something about the chronology of certain outcomes. Binaries are broken up, or SIMs occur, from the first encounters. However, we see that ejection and binary coalescence events (labelled as contact and inspiral in the figure) tend to occur later. This is not surpising; in the case of ejections, the binary must first harden via exchanges and fly-bys until the recoil is sufficient to remove it from the cluster. To produce a coalescence (labelled inspiral in the figure), the binary first passes through a CE phase, the two compact objects being left $\sim 3 R_\odot$ apart. These two stars then spiral together as angular momentum is lost through gravitational radiation on a timescale $\sim 10 - 100$Gyr. Binaries are brought into contact as angular momentum is lost via a magnetized wind from a main-sequence star component of the binary, on a timescale $\sim 1 - 100$Gyr. Single-merged objects (SMOs) are produced only for encounters involving hardened binaries hence are only seen after $\sim 10 - 100$Gyr.

In Table 4 we list the frequencies of the various possible outcomes. For the systems brought into contact by angular momentum loss from a magnetized wind we differentiate between five possible outcomes: if a main-sequence star is filling its roche lobe and transferring material to a neutron star an LMXB will be produced if the neutron star is the more massive of the two stars, otherwise the neutron star will be smothered by the remains of the main-sequence star (SN); we apply a similiar criterion for mass transfer onto a white dwarf, producing either a CV if the white dwarf is more massive, or a smothered white dwarf. This approximation is reasonable given the nature of these simulations. Finally, if both components of the binary are main-sequence stars, we tag the merged object as a blue straggler. This is another mechanism for producing blue stragglers in addition to those produced directly through collisions of two main-sequence stars during a 2+1 encounter.

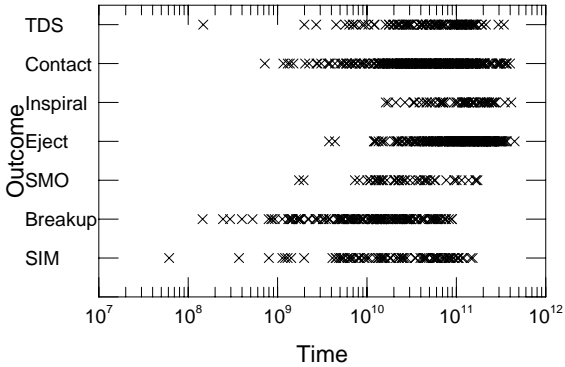

**Figure 1.** Final outcome of the binaries evolved in Run 1, as a function of when the final object was produced. SIM are scattering-induced mergers, SMOs are single merged objects, TDS refers to thick-disk systems (smothered neutron stars). Contact refers to binaries brought together via magnetic breaking, whilst inspiral refers to binaries brought together via gravitational radiation.

For systems coalescing after losing angular momentum via gravitational radiation, we distinguish between the following outcomes: the system may contain two neutron stars, a neutron star and a white dwarf, or two white dwarfs. In the latter case, we further distinguish between those systems containing a total mass greater than the Chandrasekhar limit ($1.4M_\odot$) and those less massive.

Run 2 is identical to run 1 except we that we only evolved the binaries for 15 Gyr. Of the 1000 binaries, 807 of them were unresolved after that time. In run 3, we injected the binaries at some random time drawn evenly between zero and 15 Gyr, and evolved them up to a time of 15 Gyr. In order to increase the production rate of interesting objects, we increased the number of binaries considered to 5000, of these only 507 reached some end state. Reducing the evolution time to 15 Gyr greatly reduces the number of mergers occuring as a result of gravitational radiation inspiral. In all three runs 1-3, the number of smothered neutron stars produced exceeded the number of X-ray binaries. We know that globular clusters are $\sim 15$ Gyr old, and we might expect that binaries might enter the core continuously, hence run 3 may well be the most physical of the three. In this run, we see that the number of smothered neutron stars exceeds that of X-ray binaries by a factor $\sim 4$.

In runs 4 and 5 we investigate the effect of changing the properties of the injected binaries. Runs 4 and 5 were identical to run 3, except in run 4 the masses of the two stars in the injected binaries were drawn independently from the IMF, with the extra criterion that the total mass $\geq 1.0M_\odot$ (run 4), and $\geq 0.6M_\odot$ (run 5), the latter being the mean mass of stars in the core. The binaries in runs 4 and 5 typically contain much less mass than those injected in run 3. The reduction in total mass leads to longer encounter times, and thus fewer encounters ($\sim 100$ more binaries remain unresolved in runs 4 and 5 compared to run 3). Fewer blue stragglers are produced in runs 4 and 5 (276, and 191, respectively) compared to run 3 (where the number was 355). The ratio of smothered neutron stars to X-ray binaries is reduced to $\sim 3$.

In runs 6-8, we investigate the effect of assuming smothered neutron stars in binaries behave exactly like smothered white dwarfs, with their envelopes expanding to engulf their companion and enter a common envelope phase. Run 6 is a rerun of run 5, whilst in run 7 we reduced the efficiency of the common envelope phase, $\alpha_{ce} = 0.1$, to see how much this increased the number of mergers through gravitational radiation inspiral (hereafter denoted as MGRIs), as the two stars would be left closer after a common envelope phase. In run 8, we repeated run 7, but let the binaries evolve until they reached some end state, in order to investigate the effect on the production rates of evolving the binaries for a finite time. We note that the number of MGRI events is very small in runs 6 and 7; the smothered neutron stars are not now leading to mergers of two compact objects, in 15 Gyr. We see from run 8 that if we can wait long enough, the number of MGRI events is indeed increased, and the number of neutron star/white dwarf and white dwarf/white dwarf mergers exceeds the number of X-ray binaries. However, restricting ourselves to the first 15 Gyr of the binaries' evolution, we conclude that if the smothered neutron stars do indeed evolve as assumed in runs 6-8, then the number of X-ray binaries produced will exceed the number of end-state systems containing smothered neutron stars.

### 4.2 Encounters in 47 Tucanae

We now consider encounters involving binaries in the core of the model for 47 Tuc. Runs 9-16 are exactly equivalent to runs 1-8 for $\omega$ Cen, except in the last four runs we restrict the injected binaries to systems having a total mass $> 0.7M_\odot$ (rather than the $0.6M_\odot$ used in runs 6-8), as this figure is closer to the mean mass of stars in the core.

In Figure 3 we plot the final outcomes of the 1000 binaries injected into the core in run 9. Comparison with Figure 2 clearly shows that the encounter timescale is much shorter in 47 Tuc than in $\omega$ Cen, with virtually all the binaries reaching their end states in less than 15 Gyr. As was seen in Figure 2, binaries are broken up, or undergo SIMs, before any systems are ejected from the cluster (timescale $\sim 5$-$20$ Gyr) or merge via gravitational radiation of angular momentum loss by magnetic winds (timescale $\sim 1$-$20$ Gyr).

The number of smothered neutron stars X-ray binaries produced in runs 9-16 compared to the number of blue stragglers is less than for the runs for $\omega$ Cen, owing to the relative sparsity of neutron stars in the core of 47 Tuc, compared to $\omega$ Cen. For runs 9-13 (where we halted the evolution of a binary once a smothered neutron star was produced), the number of smothered neutron stars produced exceeded the num-ber of X-ray binaries by a factor $\sim$ 4-6. The number of CVs produced greatly exceeds the number of X-ray binaries in all runs. Unlike the runs for $\omega$ Cen, the number of MGRI events is significant in all runs. If we assume white dwarf/white dwarf mergers produce neutron stars when their total mass exceeds the Chandrasekhar limit, then the number of MSPs produced through this channel may be comparable to that from smothered neutron stars.

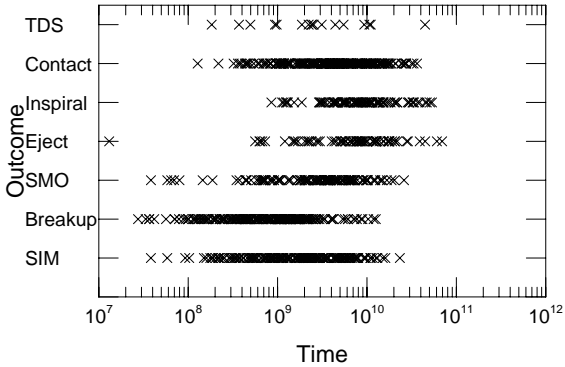

**Figure 2.** Final outcome of the binaries evolved in Run 9 (labelling same as Fig. 2).

## 5 COMPARISON OF 2+1 TO 1+1 RATES

We now compare the production rates of astrophysically interesting objects by 2+1 encounters to those from encounters between two single stars. The encounter rates listed in Table 3 are in units of encounters/$10^8$ years, hence assuming both $\omega$ Cen and 47 Tuc are 15Gyr old, we must multiply these numbers by 150 to obtain the total number of systems produced. In the runs for $\omega$ Cen we injected either 1000 or 5000 binaries into the core. For 47 Tuc, we always considered 1000 binaries. We now have to consider how many binaries we would expect to find in the two cores. The total number of stars in the cores of $\omega$ Cen and 47 Tuc are $\sim 6 \times 10^5$, and $\sim 4 \times 10^3$ respectively. Hence if we assume that $\sim 10\%$ of the stars are in binaries, then the number of binaries, $n_{\rm bin} \sim 30000$, and 2000. However if we only consider binaries containing the most massive main-sequence stars (runs 1-4, and 9-11) and assume 10% of such stars are in binaries, then the numbers are reduced to $n_{\rm bin} \sim 5000$, and 1000.

In Table 5 we list the total expected production of blue stragglers and systems containing compact objects assuming the necessary adjustment factors discussed above. We consider the rates from runs 3, 5, and 7 for $\omega$ Cen, and runs 11, 13, and 15 for 47 Tuc. Considering first the blue stragglers, we see that in $\omega$ Cen, the number produced via 2+1 encounters greatly exceeds the number produced from encounters between two single stars. By contrast in 47 Tuc, the numbers seem to be comparable for all three runs considered. Assuming blue stragglers live for $\sim 5$Gyr, we would expect $\omega$ Cen and 47 Tuc to contain $\sim 200-400$, and $\sim 400-$

500 such stars repectively. We should also note the effect of changing the masses of the binary components. In runs 3 and 11, the components of the initial binaries are taken to be turn-off mass main-sequence stars. In the other runs, they are drawn from the full range of main-sequence star masses. Comparing the number of blue stragglers produced in runs 3 and 11 to the number from the other runs (rescaled to the number of binaries injected), we see that the number is larger in runs 3 and 11. This is no surprise; we are simply injecting more massive main-sequence stars into the cluster cores.

We next consider the number of systems produced containing neutron stars. Looking first at the results for $\omega$ Cen, we see that the number of systems produced via 1+1 encounters exceed the number from 2+1 encounters (except for run 5). If we assume that all systems produced through 1+1 encounters are smothered systems, the number of smothered systems will exceed the number of X-ray binaries by a factor $\sim 2-10$. Considering now 47 Tuc, we note first that the number of systems containing neutron stars is very much less than the number of blue stragglers or systems containing white dwarfs. This is because of the relatively low number of neutron stars in the cluster model; a large fraction of those binaries producing objects of astrophysical interest have formed systems containing one of the more plentiful white dwarfs, or produced a blue straggler. We note also that the shorter encounter timescale for 47 Tuc compared to $\omega$ Cen leads to a larger number of MGRI events, as the components of post common envelope phase systems have had more time to spiral together. If we assume the merger of two white dwarfs (where the total mass exceeds the Chandrasekhar limit) produces a neutron star, such a channel will produce roughly as many neutron-star systems as 1+1 encounters, in 47 Tuc. If this is an important channel for the production of MSPs, it might help explain why 47 Tuc has been found to contain many MSPs, whilst $\omega$ Cen does not. Given an expected lifetime for an X-ray binary of $\sim 1$Gyr, we would expect to find a few such objects in $\omega$ Cen and at most one or two in 47 Tuc, assuming 1+1 encounters produce only smothered systems. Neither cluster contains a bright X-ray source, though both do contain low-luminosity binaries, which may contain accreting neutron stars (Hasinger et al. 1994, Johnston et al. 1994). The absence of bright sources in both clusters does not tell us much, as the expected number is only $\sim$ unity.

In DBH92, we concluded that no more than half the systems produced in encounters between single neutron stars and main-sequence stars would be clean binaries. If we assume half do produce LMXBs (an upper limit), then the ratio of smothered neutron stars to LMXBs will be $\sim 4 : 1$ and $\sim 5 : 1$ for $\omega$ Cen and 47 Tuc respectively. Hence even in this extreme limit, we are producing far more smothered neutron star systems than LMXBs.

Finally, we consider the systems produced containing white dwarfs. In some sense a comparison of the 1+1 to the 2+1 rates is misleading, as we have assumed in all 2+1 runs that a smothered white dwarf in a binary will essentially form a red giant and engulf its companion to form a common envelope system, and is not counted here. If we assume that none of the systems produced in 1+1 encounters produce CVs, then we conclude that the number of CVs prodcued will be much smaller than the number of isolated

**Table 5.** The total number of systems produced over 15 Gyr, via both single star/single star (1+1), and binary/single star (2+1) encounters.

| Cluster | Run | $n_{\rm bin}$ | BS | | NS | | | WD | | | MGRI | |
|---|---|---|---|---|---|---|---|---|---|---|---|---|
| | | | 1+1 | 2+1 | 1+1 | XB | SN | 1+1 | CV | SW | NW | $WD_a^2$ |
| $\omega$ Cen | 3 | 5000 | 178 | 355 | 204 | 15 | 67 | 246 | 23 | 35 | 0 | 0 |
| | 5 | 30000 | 178 | 1146 | 204 | 78 | 204 | 246 | 84 | 60 | 12 | 6 |
| | 7 | 30000 | 178 | 1008 | 204 | 72 | 0 | 246 | 42 | 18 | 6 | 12 |
| 47 Tuc | 11 | 1000 | 675 | 507 | 50 | 3 | 20 | 920 | 50 | 74 | 4 | 34 |
| | 13 | 2000 | 675 | 858 | 50 | 6 | 46 | 920 | 116 | 78 | 4 | 52 |
| | 15 | 2000 | 675 | 870 | 50 | 4 | 2 | 920 | 68 | 70 | 20 | 70 |

BS is the number of blue stragglers produced. Under NS we list the number of neutron stars involved in 1+1 encounters, the number forming X-ray binaries (XB) via 2+1 encounters, and those in smothered systems formed via 2+1 encounters. We list the equivalent three numbers for white dwarfs under WD, with CV standing for cataclysmic variables. As in Table 4, systems brought into contact via gravitational radiation are labelled MGRI (mergers through gravitational radiation inspiral). Here we subdivide this outcome into: neutron star/white dwarf mergers (NW), and white dwarf/white dwarf mergers where the total mass is greater than the Chandrasekhar mass ($WD_a^2$).

smothered white dwarfs. We expect the smothered white dwarfs to evolve to form red giants; the compact object accreting material as the envelope expands. In the case of the most-massive white dwarfs smothered by the remains of a main-sequence star, it is possible that it will accrete enough material to exceed the Chandrasekhar mass, possibly forming a neutron star via AIC. In $\omega$ Cen, the number of CVs and X-ray binaries produced seem likely to be comparable, whereas in 47 Tuc, the number of CVs will greatly exceed the number of CVs. In all six runs for the two clusters considered here, the number of white dwarf/white dwarf mergers occuring via gravitational radiation inspiral where the total mass is *less* than the Chandrasekhar limit is negligible.

## ACKNOWLEDGEMENTS

We thank Sverre Aarseth and Steinn Sigurssen for useful discussions, and the referee for their comments. MBD gratefully acknowledges support from an R.C. Tolman Research Fellowship from Caltech. The work of WB was supported in part by NSF Grant AST–0206378.

Phinney E. S. P., Kulkarni S. R., 1994, ARA&A, 32, 591
Ritter H., 1994, Mem. Soc. Astr. It., 65, 173
Sandage A., 1953, AJ, 58, 61
Verbunt F., 1987, ApJ, 312, L23
Verbunt F., Lewin W. H. G., van Paradijs J., 1989, MNRAS, 241, 51
Verbunt F., Meylan G., 1988, A&A, 203, 297
Verbunt F., Rapparport S., 1988, ApJ, 332, 193